\documentclass[epj]{svjour}
\usepackage{graphicx}
\input{epsf}
\usepackage{amssymb}

\begin{document}

\title{Poisson statistics of PageRank probabilities of 
Twitter and Wikipedia networks}

\author{
K.M. Frahm\inst{1}
\and
D.L. Shepelyansky\inst{1}
}
\institute{
Laboratoire de Physique Th\'eorique du CNRS, IRSAMC, 
Universit\'e de Toulouse, UPS, 31062 Toulouse, France
}

\titlerunning{Poisson statistics of PageRank probabilities}
\authorrunning{K.M.Frahm  and D.L.Shepelyansky}

\abstract{We use the methods of quantum chaos and Random Matrix Theory
for analysis of statistical fluctuations of PageRank probabilities
in directed networks.
In this approach the effective energy levels are
given by a logarithm of PageRank probability at a given node. 
After the standard
energy level unfolding procedure we establish that the 
nearest spacing distribution  of PageRank probabilities 
is described by the Poisson law typical for integrable quantum systems.
Our studies are done for the Twitter network and three networks of
Wikipedia editions in English, French and German.
We argue that due to absence of level repulsion the PageRank
order of nearby nodes can be easily interchanged. 
The obtained Poisson law implies that 
the nearby PageRank probabilities fluctuate as random independent variables.
}

\PACS{
{89.75.Hc}{
Networks and genealogical trees}
\and
{05.45.Mt}{
Quantum chaos; semiclassical methods}
\and
{89.75.Fb}{
Structures and organization in complex systems}
\and
{89.20.Hh}{
World Wide Web, Internet}
}

\date{Received: \today}

\maketitle

\section{Introduction}
The PageRank vector $P(K)$ of the Google matrix $G_{ij}$
had been proposed by Brin and Page 
for ranking of nodes of the World Wide Web (WWW)
in 1998 \cite{brin}.
At present the PageRank algorithm became
a fundamental element
of various search engines including Google search
\cite{meyerbook}.  This ranking works reliably  also for
other networks like the Physical Review citation network
\cite{redner,fortunato}, Wikipedia 
\cite{wikizhirov,eomepjb,eomplos} and other networks
including even the world trade network \cite{wtrade}.
Thus it is important to understand the statistical properties of the
PageRank vector.

To study the properties of PageRank probabilities
we use the standard approach \cite{brin,meyerbook}
following the notation used in \cite{eomepjb}.
The directed network
is constructed in a usual way: a directed link 
is formed from a node $j$ to a node $i$
when $j$ quotes $i$ and an element $A_{ij}$ of
the adjacency matrix is taken to be unity when
there is such a link and zero in absence of link.
Then the matrix $S_{ij}$ of Markov transitions
is constructed by normalizing elements of each column to unity
($\sum_i S_{ij}=1$) and replacing columns with only zero elements 
({\em dangling nodes}) by $1/N$, with $N$ being the matrix size.
Then the Google matrix of the network takes the form
\cite{brin,meyerbook}:
\begin{equation}
   G_{ij} = \alpha  S_{ij} + (1-\alpha)/N \;\; .
\label{eq1} 
\end{equation} 
The damping parameter $\alpha$ in the WWW context 
describes the probability 
$(1-\alpha)$ to jump to any node for a random surfer. 
For WWW the Google search engine uses 
$\alpha \approx 0.85$ \cite{meyerbook}.
The matrix $G$ belongs to the class of Perron-Frobenius 
operators \cite{meyerbook},
its largest eigenvalue 
is $\lambda = 1$ and other eigenvalues have $|\lambda| \le \alpha$. 
The right eigenvector at $\lambda = 1$, which is called the PageRank, 
has real non-negative elements $P(i)$
and gives a probability $P(i)$ to find a random surfer at site $i$. 
Thus we can  rank all nodes
in a decreasing order of PageRank probability $P(K(i))$
so that the PageRank index $K(i)$ counts all $N$ nodes
$i$ according to their ranking, placing 
the most popular nodes
at the top values $K=1,2,3 ...$.
In numerical simulations the vector $P(K_i)$
can be obtained by the power iteration method \cite{meyerbook}.
The Arnoldi method allows to compute efficiently
a significant number of eigenvalues and eigenvectors
corresponding to large values of  $|\lambda|$ 
(see e.g. \cite{arnoldibook,ulamfrahm,twitter}).

From a physical viewpoint we can make a conjecture
that the PageRank probabilities
are described by a steady-state quantum Gibbs distribution
\cite{landau}
over certain quantum levels with energies $E_i$ .
In the frame of this conjecture 
the PageRank probabilities on nodes $i$
are given by
\begin{equation}
   P(i) = \exp(-E_i/T)/Z \; , \; Z=\sum_i \exp(-E_i/T) \; 
\label{eq2} 
\end{equation} 
and inversely the effective energies  $E_i$ are given by
\begin{equation}
  E_i= - T \ln P(i) - T \ln Z\; .
\label{eq3} 
\end{equation} 
Here $Z$ is the statistical sum and $T$ is a certain effective temperature.
In some sense the above conjecture assumes that the operator matrix $G$
can be represented as a sum of two operators $G_H$ and $G_{NH}$
where $G_H$ describes a hermitian system while $G_{NH}$ represents
a non-hermitian operator
which creates a system thermalization at a certain effective temperature $T$
with the quantum Gibbs distribution over energy levels $E_i$ of operator $G_H$. 
The last term in (\ref{eq3}) 
is independent of $i$ and gives a global energy shift
which is not important. 

The statistical properties of fluctuations of levels
have been extensively studied in the fields of Random Matrix Theory (RMT)
\cite{mehta} and quantum chaos \cite{haake}.
The most direct characteristics is the probability
distribution $p(s)$ of level spacings $s$ statistics.
Here $s=(E_{i+1}-E_i)/\Delta E$ 
is a spacing between nearest levels
measured in the units of average local energy spacing
$\Delta E$. Thus the probability distribution
$p(s)$ is obtained via the unfolding procedure
which takes into account the variation of energy level density with 
energy $E$  \cite{haake}. We note that the value of $T$ in (\ref{eq3})
does not influence the statistics $p(s)$ due to spectrum unfolding
and definition of $s$ in units of local level spacing.

In the field of quantum chaos it is well established that
$p(s)$ is a powerful tool to characterize the 
spectral properties of quantum systems.
For quantum systems, which have a chaotic dynamics
in the classical limit 
(e.g. Sinai or Bunimovich billiards \cite{sinai}),
it is known that in generic cases
the statistics $p(s)$ is the same as for the RMT,
invented by Wigner to describe 
the spectra of complex nuclei \cite{mehta,bohigas,stockmann}.
This statement is known as the Bohigas-Giannoni-Schmit conjecture
\cite{bohigas}.
In such cases the distribution is well described by the so-called
Wigner surmise $p(s)=(\pi s/2) \exp(-\pi s^2/4)$
\cite{haake,stockmann}.
For integrable quantum systems (e.g. circular of elliptic billiards)
one finds a Poisson distribution $p(s)=\exp(-s)$
corresponding to the fluctuations of random independent variables.
Such a Poisson distribution  is drastically different from the RMT results
characterized by the level repulsion at small $s$ values.

The strong feature of $p(s)$ statistics is that it describes 
the universal statistical fluctuations. Thus its
use for description of PageRank fluctuations is very relevant, it
provides a new statistical information
about PageRank properties. We describe the results obtained
within such an approach in next Sections.

\section{Statistical properties of PageRank probabilities}

For our studies we use the network of entire 
Twitter 2009 studied in \cite{twitter}
with number of nodes $N=41 652 230$ and number of links 
$N_\ell=1 468 365 182 $; network of English Wikipedia
(Aug 2009; noted below as Wikipedia)
articles from \cite{wikizhirov}
with $N= 3 282 257$, $N_\ell=71 012 307$;
German Wikipedia (dated November 2013, noted below as Wikipedia-DE)
with $N=1 532 977$, $N_\ell= 36 781 077 $ and French 
Wikipedia (dated November 2013; noted below as Wikipedia-FR) 
with $N= 1 352 825$, $N_\ell=34 431 943 $. For the last two cases
we use the network data collected by S.Vigna \cite{vigna}.

\begin{figure}
\begin{center}
\includegraphics[clip=true,width=0.48\textwidth]{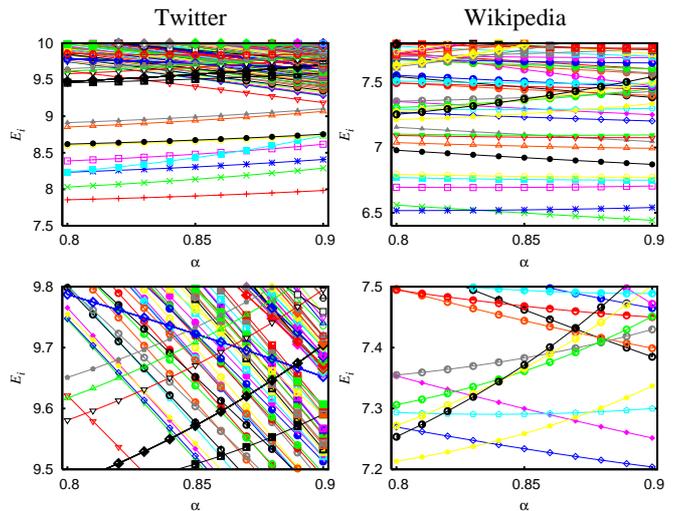}
\vglue -0.1cm
\caption{Dependence of  top PageRank levels $E_i=-\ln(P_i)$ on 
the damping factor $\alpha$ for Twitter (left panel) and Wikipedia (right 
panel). Data points on curves with the same color symbol
correspond to the same node $i$. 
The lower panels are obtained by a zoom 
in an energy range from the top panels. About 150 (for Twitter) 
or 50 (for Wikipedia) lowest levels are shown in top panels.}
\label{fig1}
\end{center}
\end{figure}

For a given network the PageRank is computed as usually by the 
power or iteration method for a
typical value of the damping factor $\alpha=0.85$.
The probabilities $P_i$ are computed with a relative
precision better than $10^{-12}$.
For each node $i$ its PageRank value $P_i$ is associated to a 
pseudo-energy 
$E_i$ by the relation $E_i=-\ln(P_i)$. 
Obviously the energy spectrum is ordered 
if the index is given in the rank index $K$, i.e. $E_{K+1}\ge E_{K}$. 
Therefore the number $n$ of levels below a given pseudo-energy 
$E$ is given by $n=K$ if $E_{K}<E<E_{K+1}$ (we also use index $i$
for $E_i$).

The evolution of energy levels $E_i$ with the variation of 
the damping factor $\alpha$
are shown in Fig.~\ref{fig1}  for Twitter and Wikipedia networks.
The results show many level crossings which are 
typical of Poisson statistics. We note that here each level
has its own index so that it is rather easy to see 
if there is a real or avoided level crossing.
In this respect the situation is simpler compared to
energy levels in quantum systems.

\begin{figure}
\begin{center}
\includegraphics[clip=true,width=0.48\textwidth]{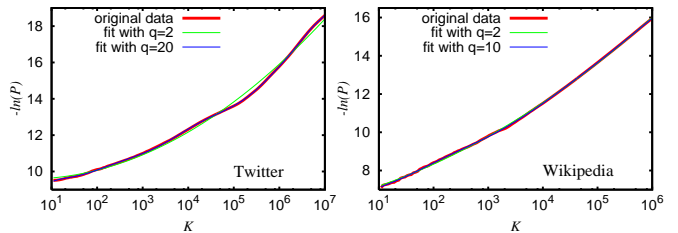}
\vglue -0.1cm
\caption{The thick red curve shows $-\ln(P)=E$ versus $K$ 
for the PageRank probability $P$ of 
Twitter (Wikipedia) in the left (right) panel. 
The thin green curve corresponds to the fit $-\ln(P)=Q(\ln(K))$ where $Q(x)$ is 
a polynomial of degree $q=2$. The thin blue curve corresponds to the fit with 
a polynomial of degree $q=20$ ($q=10$). 
The fits are obtained for the range $10<K\le 10^7$ ($10<K\le 10^6$) with 
weights $\sim 1/K$ attributed to each data point.  Here and in next Figs. $\alpha=0.85$.
}
\label{fig2}
\end{center}
\end{figure}

In the following we fix the damping factor to the standard value 
$\alpha=0.85$. To obtain the unfolded spectrum 
with an average uniform level spacing of unity 
(see e.g. \cite{haake}) 
one has to replace the 
function $E_i$ by a smooth function. As shown in Fig.~\ref{fig2}, one 
can very well approximate $E_K$ by a polynomial $Q(x)$ of modest degree 
in the variable $x=\ln(K)$. In this procedure
it is better to exclude the first ten nodes with $K\le 10$ 
which do not affect the global statistics.
For a fit range $10<K\le 10^4$ a polynomial of 
degree 2 is already sufficient. However, for larger intervals, e.g. 
$10<K\le 10^7$ for Twitter or 
$10<K\le 10^6$ for Wikipedia it is better to increase the polynomial degree 
up to 20. Once the polynomial fit is known one obtains the unfolded 
energy eigenvalues $S_i$ by solving the equation $E_i=Q(\ln(S_i))$ 
using the Newton method. For each energy the obtained value of 
$S_i \approx i$ is 
rather close to $K=i$ index with an average spacing of unity. 
In certain cases this equation does not provide a 
solution for energies close to the boundary of the fit range. In these cases 
the unfolded spectrum is slightly reduced with respect to the initial fit 
range. 

\begin{figure}
\begin{center}
\includegraphics[clip=true,width=0.48\textwidth]{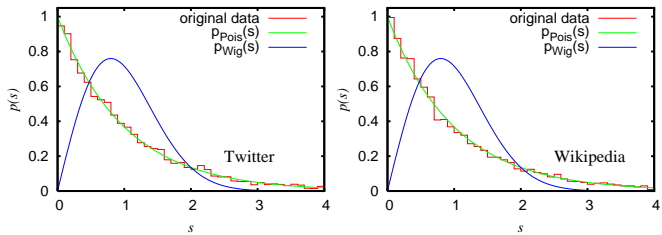}
\vglue -0.1cm
\caption{Histogram of unfolded level spacing statistics 
using pseudo-energies $E_i=-\ln(P_i)$ of Twitter (Wikipedia) shown 
in the left (right) panel. The unfolding is done with the 
fit shown in Fig. \ref{fig2} using a polynomial of degree 2 and a fit 
range $10<K\le 10^4$. The Poisson distribution 
$p_{\rm Pois}(s)=\exp(-s)$ 
and the Wigner surmise 
$p_{\rm Wig}(s)=\frac{\pi}{2}\,s\,\exp(-\frac{\pi}{4}\,s^2)$ 
are also shown for comparison. }
\label{fig3}
\end{center}
\end{figure}

In Fig.~\ref{fig3} only a polynomial of degree 2 is used since the fit range 
$10<K\le 10^4$ is rather small and the histogram fluctuations, compared 
with the Poisson distribution, are still quite considerable due to the limited 
number of $N_s \sim 10^4$ data points. The obtained data
show a good agreement of  results  with the Poisson statistics.

\begin{figure}
\begin{center}
\includegraphics[clip=true,width=0.48\textwidth]{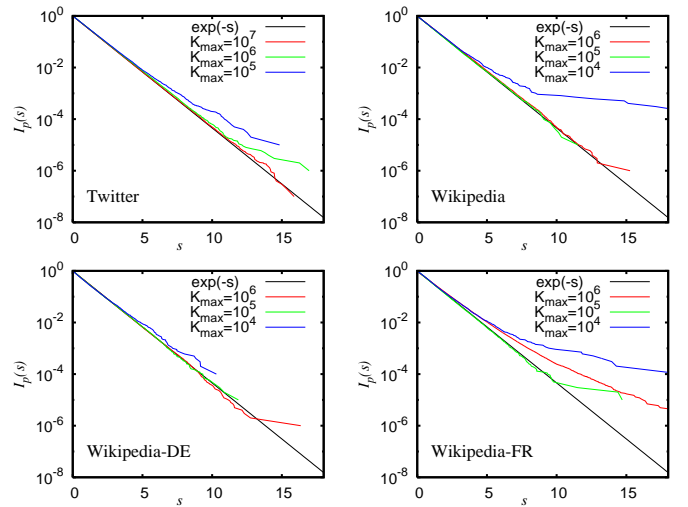}
\vglue -0.1cm
\caption{The color curves show the integrated probability 
$I_p(s)=\int_s^\infty d\tilde s\,p(\tilde s)$, given in semi-logarithmic 
representation,  for the 
PageRank probabilities 
for networks of Twitter, Wikipedia, Wikipedia-DE and Wikipedia-FR. 
The unfolding is done as in Fig.~\ref{fig2} using 
a fit polynomial of degree 20 and a fit range $10<K\le K_{\rm max}$ 
with three different values of $K_{\rm max}$ given in the panels. The black 
line corresponds to $I_p(s)=\exp(-s)$ obtained for the case of Poisson 
distributed levels. 
}
\label{fig4}
\end{center}
\end{figure}

In Fig.~\ref{fig4} we show the integrated probability
to find a level spacing larger than $s$:
\begin{equation}
\label{eq4}
I_p(s)=\int_s^\infty d\tilde s\,p(\tilde s) \; .
\end{equation}
This quantity is numerically more stable 
since no histogram is required. One simply 
orders the spacings $s_i=S_{i+1}-S_i$ and draws the ratio 
$1-i/N_s$ versus $s_i$ where $i$ is the ordering index of the spacings 
and $N_s$ is the number of spacings in the numerical data. 

The data shown in  Fig.~\ref{fig4} clearly demonstrate 
that $I_p(s)$ follows the 
Poisson expression $I_p(s)=\exp(-s)$ for 
a quite large range of level spacings. 
Of course, for the largest values of $s$ there are deviations which are 
either due to the lack of statistics (especially for modest values of 
$K_{\rm max}$) or due to the fact that the number of levels is close to 
the total network size. 

We also note that for large
values of $K \ge 10^6$ there are $N_d$ degenerate nodes with 
identical $P(i)$ values with at least one more another node or a few nodes. 
Such an effect has been pointed in \cite{twitter}.
These artificial degeneracies provide an additional delta function 
contribution 
$w_0\,\delta(s)$ in the Poisson statistics $p(s)$ where $w_0$ is the 
probability to find such a degeneracy. 
There are about $N_d \approx 10^2$ ($N_d\approx 10^5$) degeneracies 
for Twitter nodes for $K<10^6$ ($K<10^7$) 
which gives $w_0\approx 10^{-4}$ ($w_0\approx 10^{-2}$). In a histogram 
of bin-width $\Delta s=0.1$ this gives a relative change of the height of 
the first bin 
at $s=0$ of $10\,w_0\approx 10^{-3}$ ($\approx 10^{-1}$) and unless 
we use too large $K$ value the statistical contribution of such degenerate 
nodes is indeed very small. 

We note that if we use all nodes of Twitter up to $K<4.2\times 10^7$ we have 
$N_d\approx 1.1\times 10^7$ with $w_0\approx 0.26$ which is indeed 
considerable. In this particular case also the distribution of close 
degeneracies ($0<s\ll 1$) is quite different from 
the (rescaled) Poisson distribution $(1-w_0)\,\exp[-(1-w_0)\,s]$ 
for the non-degenerate levels. Apparently a particular network structure,
which is responsible for the degeneracies, also enhances the number of 
close degeneracies.  We attribute the appearance of such degeneracies 
to a weak interconnections between nodes at the tail of
PageRank probability where the fluctuations are not stabilized
being sensible to
the finite network size.

\medskip
\begin{table}\label{table1}
\centering
\begin{tabular}{|l|l|l|l|l|}
\hline
\ K\  &\  $S_i-S_{i-1}$\ &\ $S_{i+1}-S_{i}$\  &\ Title\ \\
\hline
\hline
\ 996\ &\  8.43535\ &\ 6.57294\ &\ Henry VIII of England\ \\
\hline
\ 2966\ &\  4.07317\ &\ 4.09474\ &\ The Age\ \\
\hline
\ 3398\ &\  4.21163\ &\ 4.65018\ &\ Debt\ \\
\hline
\ 3982\ &\  4.30229\ &\ 4.01818\ &\ GREEN\ \\
\hline
\ 6098\ &\  4.42446\ &\ 4.78164\ &\ Vomiting\ \\
\hline
\ 6632\ &\  4.22776\ &\ 4.38045\ &\ Mary I of Scotland\ \\
\hline
\ 9388\ &\ 4.42904\ &\ 4.94249\ &\ Simulation\ \\
\hline
\end{tabular}
\caption{List of nodes with unfolded neighbor level spacings 
$s_i=S_i - S_{i-1} > 4$ for Wikipedia network.}
\end{table}

Our data show that the Poisson statistics gives a good description
of fluctuations of PageRank probabilities. It may be interesting to
determine what are the nodes
which have very large spacings $s$ from nearest levels
on both sides. It is natural to expect that those nodes
will be rather stable in respect  to modifications
of network or damping factor variations.
Such nodes for Wikipedia network are shown in 
Table 1
for $s>4$ and $K <10^4$.
Such a selection captures two important figures of English history
but the reasons for appearing of other nodes still need to be clarified.
We think that a further study of nodes with large statistical deviations
of spacing values can provide a new interesting information about robust
nodes of a given network. The validity of the Poisson statistics
means that the ranking order can be easily
interchanged between nodes with nearby values of PageRank index $K$.

We also analyzed the statistics of PageRank probabilities for a
random triangular matrix model (triangular RPFM) 
introduced in \cite{physrev}. We find here the Poisson statistics. We also 
consider CheiRank probability vector of Wikipedia
(it is given by the PageRank probability
for the Wikipedia network with inverted direction of links)
\cite{wikizhirov} and also find here the Poisson distribution.

\section{Discussion}

We use the methods of quantum chaos to study 
the statistical fluctuations of PageRank
probabilities in four networks 
of Twitter, Wikipedia English, German and French.
We associated the effective pseudo-energy levels $E_i$
to PageRank probabilities via the relation
$E_i = - \ln P_i$ and use the unfolding 
level density procedure to have
homogeneous spacings between levels.
This procedure is commonly used in
the field of quantum chaos (see e.g. \cite{haake,stockmann}).
Our studies show that the level spacing statistics is
well described by the Poisson distribution $p(s)=\exp(-s)$.
Thus there is any sign of level repulsion typical
of the quantum chaotic billiards \cite{bohigas}
and RMT \cite{mehta}. Such a result can be considered as 
a natural one for nodes with large values of PageRank index $K$
where nodes can be assumed as independent. However, 
the Poisson distribution
remains valid even for relatively low values
$K \leq 10^4$ where a significant number of
links exist between the users of Twitter
as discussed in \cite{twitter}. Thus even a large number
of links between top nodes does not lead 
to their interdependence so that
nearby PagRank probabilities behave themselves 
as random independent variables.

We should note that the relation $E_i = - \ln P_i$,
used in our studies to have a correspondence with level 
spacing statistics, is not really so important since after that we
apply the unfolding procedure.
Due to this our method simply gives us the fluctuations of nearby
PageRank probabilities in a correctly weighted 
dimensionless representation where the validity
of Poisson distribution becomes directly visible.
We think that the investigation of nodes with large 
spacings with nearby nodes in $K$ can provide
a new useful information
for network analysis.



\section*{Acknowledgments}

This research is supported in part by the EC FET Open project
``New tools and algorithms for directed network analysis''
(NADINE $No$ 288956). We thank Sebastiano Vigna for
providing us the network data
for German and French Wikipedia,
collected in the frame of NADINE project;
these data sets can be obtained from the web page
of S.Vigna \cite{vigna}.

\end{document}